\documentstyle[psfig]{l-aa}
\input epsf
\begin{document}
\thesaurus{11                
	      (11.01.2;      
	       11.14.1;      
	       11.19.1;      
 	       11.02.1;      
	       11.17.3;      
	       11.19.03)}    
\title{
Miscellaneous observations of active galactic nuclei. II.
\thanks{Based on observations collected at the Observatoire de Haute-Provence 
(CNRS), France, and at the European Southern Observatory, La Silla, Chili.}}
\subtitle{ }
\author{A. C. Gon\c{c}alves, P. V\'eron and M.-P. V\'eron-Cetty}
\offprints{P. V\'eron}
\institute{Observatoire de Haute Provence (CNRS), F-04870 Saint Michel
l'Observatoire, France}
\date{Received 20 March 1997 / Accepted 22 April 1997 }
\maketitle


\begin{abstract}

We observed 37 AGN candidates and classified them on the basis of their 
spectroscopic properties; three are 
confirmed QSOs, one is a BL Lac object, nine are Seyfert 1 
galaxies, four Seyfert 2s, while twenty are HII regions.\\
\keywords{Galaxies: active -- Galaxies: nuclei -- Galaxies: Seyfert -- 
Galaxies: BL Lacertae objects: general -- Galaxies: quasars: general -- 
Galaxies: starburst} 
\end{abstract}

\section{Introduction}

In the course of several observing runs, we obtained optical spectra 
of 37 AGN candidates with uncertain classification. Twenty of them 
turned out to be extragalactic HII regions ionized by hot stars, while 
seventeen were confirmed to be QSOs, BL Lac or Seyfert galaxies. In a previous 
paper (V\'eron-Cetty \& V\'eron, 1986), the classification of 61 AGN 
candidates was given.

\section{Observations}

Most of the observations were carried out in 1995, 1996 and 1997 
with the spectrograph CARELEC (Lema\^{\i}tre et al., 1989) attached to the 
Cassegrain focus of the Observatoire de Haute Provence (OHP) 1.93m teles\-cope. 
The spectrograph settings used during these runs are given in Table 1.
The detector was a $512\times512$ pixels, $27\times27\/\mu$m  Tektronic CCD. 
The slit width was 2.1 arcsec, corresponding to a projected slit width on the 
detector of 52~$\mu$m, or 1.9 pixel. The resolution, as measured on the night 
sky emission lines, was 13.5 and 3.4 \AA\, FWHM at low and high resolution, 
respectively. The spectra were flux calibrated using the standard stars 
given in Table 1, taken from Oke (1974), 
Stone (1977), Oke \& Gunn (1983) or Massey et al. (1988). 

Additional observations were made with EFOSC (Dekker et al., 1988) at 
the 3.6m ESO telescope in La Silla during two runs in July 1995 and 
August 1996. 
The detector was the CCD ESO\#26, similar to the one used at OHP. The 
dispersion was 230 \AA mm$^{-1}$; the slit width was 1.5 arcsec, corresponding 
to 2.2 pixels and resulting in a resolution of 15\AA. The wavelength range was 
$\lambda\lambda3500-7600$ \AA. The spectra were flux calibrated using 
the standard stars W~485A and GD~190 (Oke, 1974).

\begin{table}[h]
\begin{center}
\begin{tabular}{p{2.3cm}rcl}
\hline
Date & Dispersion & $\lambda$ Range & \verb+  +Standard \\
 & ($\rm \AA$ mm$^{-1}$) & ($\rm \AA$) & \verb+    +stars\\
\hline
21 -- 23.03.95 & 66\verb+   + & 6500 -- 7400 & \verb+ +BD~26$\rm ^{o}$2606 \\
24 -- 28.08.95 & 260\verb+   + & 4500 -- 8000 & \verb+ +BD~25$\rm ^{o}$3941 \\ 
 & & & \verb+ +BD~28$\rm ^{o}$4211 \\
28 -- 31.08.95 & 66\verb+   + & 6700 -- 7600 & \verb+ +BD~25$\rm ^{o}$3941 \\
 & & & \verb+ +BD~28$\rm ^{o}$4211 \\
31.08 -- 04.09.95 & 66\verb+   + & 4860 -- 5760 & \verb+ +Feige 15 \\
 & & & \verb+ +BD~28$\rm ^{o}$4211 \\
10.05.96 & 66\verb+   + & 6700 -- 7600 & \verb+ +GD~140 \\
 & & & \verb+ +BD~26$\rm ^{o}$2606\\
13.05.96 & 66\verb+   + & 4860 -- 5760 & \verb+ +Feige~98 \\
 & & & \verb+ +Kopff 27\\
08.06.96 & 66\verb+   + & 4860 -- 5760 & \verb+ +Feige~66 \\
 & & & \verb+ +Kopff 27\\
09.06.96 & 66\verb+   + & 6700 -- 7600 & \verb+ +Feige~66 \\
 & & & \verb+ +BD~28$\rm ^{o}$4211\\
06 -- 08.01.97 & 66\verb+   + & 4720 -- 5620 & \verb+ +EG~247 \\
08 -- 11.01.97 & 66\verb+   + & 6175 -- 7075 & \verb+ +EG~247 \\
\hline
\end{tabular}
\caption{Spectrograph settings and standard stars during OHP observations.}
\end{center}
\end{table}

\begin{table*}[t]
\begin{center}
\begin{tabular}{p{3cm}cccclccc}
\hline
Name & Disp. & Date & Exp. time & $\;\;$ & Name & Disp. & Date & Exp. time \\
 & & & (min) & & & & & (min) \\
\hline
4C 12.05 & D & 10.08.96 & $10\:\;$ & & 
PKS $1420-27$ & D & 25.07.95 & $10\:\;$ \\
Mark 1147 & A & 26.08.95 & $20\:\;$ & &
Mark 816 & B & 13.05.96 & $20\:\;$ \\
Mark  971 & B & 01.09.95 & $20\:\;$
& &PKS $1437-153$ & D & 27.07.95 & $10\:\;$ \\
Mark  998 & A & 26.08.95 & $20\:\;$ 
& &Mark 833 & C & 22.03.95 & $20\:\;$ \\
Q $0155+0220$ & B & 31.08.95 & $20\:\;$ & & 
Mark 483 & C & 22.03.95 & $20\:\;$ \\
Mark 596 & A & 25.08.95 & $20\:\;$ & &
KUV $15519+2144$ & C & 22.03.95 & $20\:\;$ \\
KUV $03079-0101$ & A & 27.08.95 & $20\:\;$ & &
Q $1619+3752$ & B & 01.09.95 & $20\:\;$ \\
CBS 74 & A & 30.04.95 & $15\:\;$ & &
EXO $1622.0+2611$ & C & 22.03.95 & $20\:\;$ \\
 & C & 22.03.95 & $20\:\;$ & &
Q $1624+4628$ & C & 28.08.95 & $20\:\;$ \\
HS $0843+2533$ & C & 10.01.97 & $20\:\;$ & &
Q $1638+4634$ & A & 27.08.95 & $20\:\;$ \\
Mark 391 & B & 07.01.97 & $20\:\;$ & &
Kaz 110 & B & 23.06.96 & $20\:\;$ \\
 & C & 10.01.97 & $20\:\;$ & &
 & C & 22.06.96 & $20\:\;$ \\
KUG $0929+324$ & C & 21.03.95 & $20\:\;$ & &
PKS $1903-80$ & D & 28.07.95 & $10\:\;$ \\
CG 49 & C & 10.05.96 & $20\:\;$ & &
RN 73 & B & 31.08.95 & $20\:\;$ \\
UM 446 & C & 21.03.95 & $20\:\;$ & &
 & C & 30.08.95 & $20\:\;$ \\
US 2896 & C & 22.03.95 & $20\:\;$ & &
Q $2233+0123$ & C & 30.08.95 & $20\:\;$ \\ 
Mark 646 & C & 22.03.95 & $20\:\;$ & &
Q $2257+0221$ & B & 02.09.95 & $20\:\;$ \\
2E $1219+0447$ & C & 22.03.95 & $20\:\;$ & &
NGC 7678 & B & 02.09.95 & $20^{*}$ \\
KUV $13000+2908$ & C & 22.03.95 & $20\:\;$ & &
 & C & 30.08.95 & $20\:\;$ \\
Q $1356-067$ & C & 22.03.95 & $20\:\;$ & &
E $2344+184$ & B & 01.09.95 & $20\:\;$ \\
Mark 469 & B & 08.06.96 & $20\:\;$ & &
UM 11 & B & 02.09.95 & $20^{*}$ \\
 & C & 09.06.96 & $20\:\;$ & &
  & C & 29.08.95 & $20\:\;$ \\
\hline
\end{tabular}
\caption{Journal of observations. 
A: OHP, 260 \AA mm$^{-1}$; 
B: OHP, 66 \AA mm$^{-1}$ blue; 
C: OHP, 66 \AA mm$^{-1}$ red; 
D: ESO, 230 \AA mm$^{-1}$.
An `` * '' after the exposure time indicates the presence of 
clouds during the exposure.}
\end{center}
\end{table*}

The journal of observations is given in Table 2 and the 
list of the observed objects with relevant data, in Table 3. 
The spectra were analysed 
in terms of Gaussian components as described in V\'eron et al. (1997). 
Table 4 gives for each object 
the velocity, width and relative strength of each line, together with the 
adopted classification. Objects with broad Balmer lines were classified 
as Seyfert 1 galaxies, or QSOs whenever their absolute magnitude was brighter 
than $\rm M_{B} = -23.0$  (assuming H$_{0}$ = 50 kms$^{-1}\/ \rm Mpc^{-1}$); 
Seyfert 2s and HII regions were distinguished on 
the basis of the value of the $\lbrack \rm OIII \rbrack \lambda5007 / H\beta$ 
and $\lbrack \rm NII \rbrack \lambda6584 / H\alpha$ line ratios (Veilleux 
and Osterbrock, 1987). In some cases, the classification is based on a single 
line ratio, either $\lambda$5007/H$\beta$ or $\lambda$6584/H$\alpha$. This is
potentially dangerous; however we think that in most cases, there is no 
ambiguity, specially when $\lambda$6584/H$\alpha$ $<$ 0.3 (see Fig. 6a in 
V\'eron et al., 1997).  

\section{Notes on individual objects}

{\bf 4C 12.05} (Gower et al., 1967) = PKS 0035+121 (Shimmins et al., 
1975) has been tentatively identified by Wills \& Wills (1976) and 
Jauncey et al. (1978) with a 16.5-17.0 mag object, the position of which 
is in good agreement with the accurate radio position measured by Condon 
et al. (1977); but they have shown that the optical spectrum, although 
inconclusive, was probably that of a star. However, Wills \& Wills have 
remarked that there appears to be a small, fainter, south-preceeding blue 
object, visible on the Palomar Sky Survey prints, blended with the 
image of the star. A V image, obtained on August 10, 1996 with EFOSC 
at the 3.6m ESO telescope on La Silla shows that, indeed, the object is 
double, with a separation of 2.8 arcsec. The spectrograph slit was aligned 
on the two objects (PA=$229\rm^{o}$); a 10 min exposure spectrum shows 
the north-following object to be a star, while the south-preceeding object 
is a QSO at z=1.395 (Fig. 1). The magnitude of the QSO, as measured on 
the spectrum 
is about 0.55 mag weaker than the star in B, and 0.87 mag in V. The emission
line fluxes are 570 and 360 $10^{-16}\, \rm erg\,s^{-1}\,cm^{-2}$ for 
$\rm CIII \rbrack\lambda1909$ and $\rm MgII\,\lambda2798$ respectively. 


{\bf Mark 1147} is an emission line galaxy (Markarian et al., 
1980); it has been erroneously classified as a Seyfert 1 by 
V\'eron-Cetty \& V\'eron (1985). Our low dispersion spectrum (Fig. 2) 
shows that it is a HII region, with 
$ \rm \lambda5007/H\beta = 2.18$ and $ \rm \lambda6584/H\alpha = 0.24$; this 
is in agreement with Markarian et al. (1980), who have noticed that 
$\lambda6584$ is weak compared to H$\alpha$.

{\bf Mark 971} = KUG 0101+353 (Takase \& Miyauchi-Isobe, 1991b). Markarian 
et al. (1984) suggested that it could have an active nucleus; this, however, 
was not confirmed neither by Denisyuk \& 
Lipovetski (1984) nor by Lipovetski et al. (1989). 
Our spectrum (Fig. 6) shows narrow 
emission lines ($<280$ kms$^{-1}$ FWHM) with 
$ \rm \lambda5007/H\beta = 0.41$ together with an H$\beta$ line in 
absorption; this object is, therefore, a HII region.

\begin{table*}[t]
\begin{center}
\begin{tabular}{p{3cm}ccclccc}
\hline
Name & $\alpha$ & $\delta$ & z & mag & old class. & our class. & Ref. \\
\hline
4C 12.05 & 00\/ 35\/ 41.98\/ & $\; $ 12\/ 11\/ 03.6\/ $\: $ &
-- & 17.5 & ? & Q & (25) \\
Mark$\,1147$ & 00\/ 45\/ 57.94\/ & $\; $ 10\/ 03\/ 56.9\/ $\: $ & 
0.036 & 15.7 & S1 & HII & (18) \\ 
Mark$\;\;971$ & 01\/ 01\/ 32.79\/ & $\; $ 35\/ 18\/ 07.8\/ $\: $ & 
0.085 & 16.5 & S? & HII & (16) \\
Mark$\;\;998$ & 01\/ 30\/ 02.04\/ & $\/ -\,02$\/ 20\/ 11.9\/ $\;\; $ & 
0.078 & 16.0 & S? & HII & (17) \\
Q $0155+0220$ & 01\/ 55\/ 47.90\/ & $\; $ 02\/ 20\/ 26.3\/ $\: $ & 
0.066 & 16.7 & ? & HII & (23) \\
Mark$\;\;596$ & 02\/ 40\/ 12.67\/ & $\; $ 07\/ 23\/ 09.5$\/$* & 
0.038 & 14.8 & S? & S2 & (13) \\
KUV $03079-0101$ & 03\/ 07\/ 54.83\/ & $\/ -\,01$\/ 01\/ 10.3\/ $\;\; $ & 
0.080 & 16.3 & ? & S1.0 & (7) \\
CBS 74 & 08\/ 29\/ 11.46\/ & $\; $ 37\/ 17\/ 49.1\/ $\: $ & 
0.091 & $17.\;\; $ & S & S1.2 & (21) \\ 
HS $0843+2533$ & 08\/ 43\/ 56.47\/ & $\; $ 25\/ 33\/ 14.7\/ $\: $ & 
0.050 & 16.8 & S1 & S1 & - \\
Mark$\;\;391$ & 08\/ 51\/ 32.38\/ & $\; $ 39\/ 43\/ 45.5\/ $\: $ &
0.013 & 14.1 & S? & HII & (11) \\ 
KUG $0929+324$ & 09\/ 29\/ 01.90\/ & $\; $ 32\/ 26\/ 59.9$\/$* & 
0.005 & 17.5 & ? & HII & (24) \\
CG 49 & 09\/ 58\/ 07.76 & $\; $ 31\/ 26\/ 44.7\/ $\: $ &
0.042 & 16.4 & S2 & S2 & (20) \\
UM 446 & 11\/ 39\/ 12.06\/ & $\/ -\,01$\/ 37\/ 27.1\/ $\;\; $ & 
0.005 & 17.3 & ? & HII & (8) \\
US 2896 & 11\/ 42\/ 33.73\/ & $\; $ 31\/ 03\/ 56.5\/ $\: $ & 
0.060 & 16.0 & S1.5 & S1 & (22) \\
Mark$\;\;646$ & 12\/ 03\/ 17.01\/ & $\; $ 35\/ 27\/ 27.5\/ $\: $ & 
0.054 & 15.3 & S & S1 & (14) \\
2E $1219+0447$ & 12\/ 19\/ 04.62\/ & $\; $ 04\/ 47\/ 04.3\/ $\: $ & 
0.094 & 16.8 & ? & S1 & (3) \\
KUV $13000+2908$ & 13\/ 00\/ 01.52\/ & $\; $ 29\/ 07\/ 37.0\/ $\: $ & 
0.023 & 16.1 & S2 & HII & (7) \\
Q $1356-067$ & 13\/ 56\/ 44.90\/ & $\/ -\,06$\/ 07\/ 43.8\/ $\;\; $ & 
0.072 & 16.2 & S? & HII & - \\
Mark$\;\;469$ & 14\/ 16\/ 12.98\/ & $\; $ 34\/ 35\/ 46.0\/ $\: $ &
0.069 & 16.0 & ? & HII & (12) \\
PKS $1420-27$ & 14\/ 19\/ 55.50\/ & $\/ -\,27$\/ 14\/ 20.8\/ $\;\; $ & 
-- & $18.\;\; $ & Q? & Q & (2) \\
Mark$\;\;816$ & 14\/ 31\/ 40.78\/ & $\; $ 52\/ 59\/ 26.8\/ $\: $ &
0.089 & 16.5 & S? & HII & (15) \\
PKS $1437-153$ & 14\/ 37\/ 11.31\/ & $\/ -\,15$\/ 18\/ 58.9\/ $\;\; $ & 
-- & $19.\;\; $ & Q? & BL & (4) \\
Mark$\;\;833$ & 14\/ 55\/ 59.69\/ & $\; $ 35\/ 24\/ 05.4\/ $\: $ & 
0.040 & 16.0 & S? & HII & (15) \\
Mark$\;\;483$ & 15\/ 28\/ 41.48\/ & $\; $ 34\/ 05\/ 53.3\/ $\: $ & 
0.048 & 16.4 & HII & HII & (12) \\
KUV $15519+2144$ & 15\/ 51\/ 53.33\/ & $\; $ 21\/ 43\/ 42.9\/ $\: $ & 
0.040 & 15.8 & S2 & HII & (7) \\
Q $1619+3752$ & 16\/ 19\/ 55.81\/ & $\; $ 37\/ 52\/ 36.3\/ $\: $ & 
0.034 & 17.3 & ? & HII & (23) \\
EXO $1622.0+2611$ & 16\/ 22\/ 05.33\/ & $\; $ 26\/ 11\/ 23.9\/ $\: $ & 
-- & 16.1 & S? & S1 & - \\
Q $1624+4628$ & 16\/ 24\/ 34.76\/ & $\; $ 46\/ 28\/ 48.2\/ $\: $ & 
0.030 & 16.1 & ? & HII & (23) \\
Q $1638+4634$ & 16\/ 38\/ 50.52\/ & $\; $ 46\/ 34\/ 38.8\/ $\: $ & 
0.059 & 16.4 & ? & HII & (23) \\
Kaz 110 & 16\/ 57\/ 16.83\/ & $\; $ 69\/ 09\/ 08.0\/ $\: $ & 
0.053 & 17.2 & HII & HII & (5) \\
PKS $1903-80$ & 19\/ 03\/ 56.16\/ & $\/ -\,80$\/ 14\/ 59.8\/ $\;\; $ & 
-- & 19.0 & Q? & Q & (1) \\
RN 73 & 20\/ 36\/ 08.47\/ & $\; $ 88\/ 02\/ 05.4\/ $\: $ & 
0.047 & 17.5 & ? & S1.9 & (19) \\
Q $2233+0123$ & 22\/ 33\/ 08.78\/ & $\; $ 01\/ 24\/ 00.2\/ $\: $ & 
0.058 & 16.6 & ? & S1 & (23) \\
Q $2257+0221$ & 22\/ 57\/ 00.37\/ & $\; $ 02\/ 21\/ 29.8\/ $\: $ & 
0.048 & 16.7 & ? & S2 & (23) \\
NGC 7678 & 23\/ 25\/ 57.91\/ & $\; $ 22 08 44.7\/ $\: $ & 
0.012 & 15.3 & S2 & HII & (6) \\
E $2344+184$ & 23\/ 44\/ 53.30\/ & $\; $ 18 28 10.8\/ $\: $ & 
0.138 & 15.9 & ? & S2 & (10) \\
UM 11 & 23\/ 50\/ 45.22\/ & $\; $ 03 26 22.3$\/$* & 
0.038 & 16.0 & S & HII & (9) \\
\hline
\end{tabular}
\caption{This table gives for each of the observed objects: the name (col. 1), 
the B1950 optical position measured on the 
Digitized Sky Survey (col. 2 and 3), where the the r.m.s. error is 
0.6~arcsec in each coordinate; `` * '' indicates objects with 
larger errors because of their location near one edge of the 
Schmidt plate (V\'eron-Cetty \& V\'eron, 1996), 
the published redshift (col. 4), the magnitude (col. 5), the old spectral 
classification: S2: Seyfert 2, S1: Seyfert 1, HII: HII region, Q: Quasar, 
?: unknown (col. 6), and our classification (col. 7). 
References for the finding charts (col. 8): (1) Anguita et al. (1979), 
(2) Bolton \& Ekers (1966), (3) Bowen et al. (1994), (4) Condon et al. (1977), 
(5) Kazarian (1979), (6) Kazarian \& Kazarian (1980), (7) Kondo et al. (1984), 
(8) MacAlpine \& Williams (1981), (9) MacAlpine et al. (1977), 
(10) Margon et al. (1985), (11) Markarian \& Lipovetski (1971),
(12) Markarian \& Lipovetski (1972), 
(13) Markarian \& Lipovetski (1973), (14) Markarian \& Lipovetski (1974), 
(15) Markarian \& Lipovetski (1976), (16) Markarian et al. (1977a), 
(17) Markarian et al. (1977b), (18) Markarian et al. (1979), 
(19) Penston (1971), 
(20) Pesch \& Sanduleak (1983), 
(21) Pesch \& Sanduleak (1986), (22) Sanduleak \& Pesch (1984), 
(23) Schneider et al. (1994), (24) Takase \& Miyauchi-Isobe (1985), 
(25) Wills \& Wills (1976).}
\end{center}
\end{table*}

{\bf Mark 998}. According to Markarian et al. (1984), this galaxy could have 
an active nucleus; Denisyuk \& Lipovetski (1984) and Lipovetski et al. (1989) 
were not able to confirm this. Our low dispersion spectrum (Fig. 2) 
shows $ \lambda6584/\rm H\alpha = 0.23 $. 
It follows that this object is most probably a HII region.  

{\bf Q 0155+0220} is an emission line galaxy according to Schneider et al. 
(1994). Our spectrum (Fig. 6) shows it to be a HII 
region with narrow (FWHM $< 325$ kms$^{-1}$) H$\beta$ and 
$\lambda5007$ emission lines, and $ \lambda5007/\rm H\beta = 0.71 $.

\begin{figure*}[t]
\epsfxsize=18cm \epsfysize=18.7cm \epsfbox{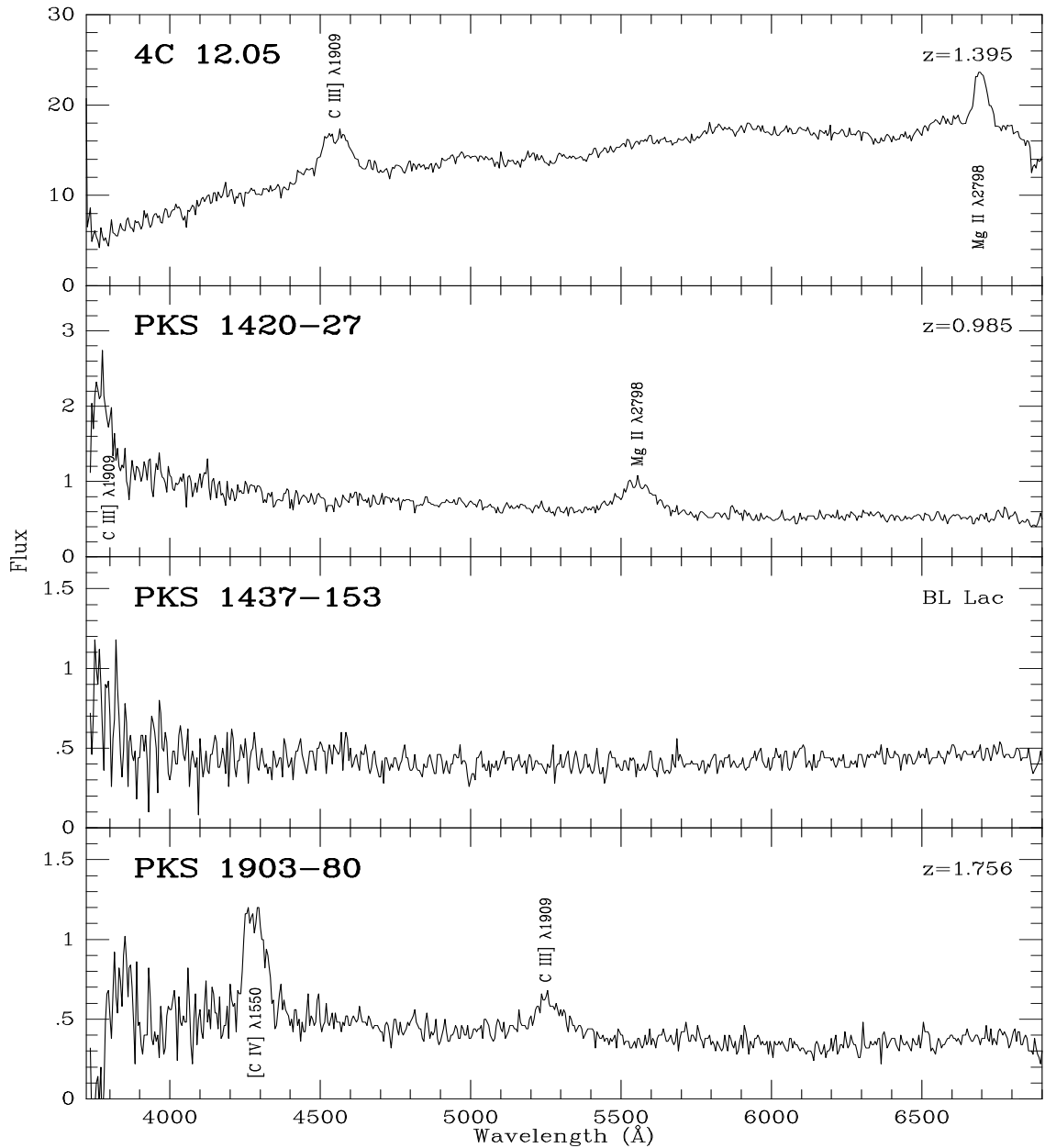}
\caption{Low dispersion spectra (resolution $\sim 15$ \AA) of four objects
observed with the 3.6m ESO telescope. The fluxes are in units of
$\rm 10^{-16}\,erg\,s^{-1}cm^{-2}\AA^{-1}$.}
\label{QSR}
\end{figure*}

{\bf Mark 596}. This object, having $ \lambda6584/\rm H\alpha > 1 $ may 
have an active nucleus (Markarian et al., 1984). It is indeed a Seyfert 2 
galaxy as our spectrum (Fig. 2) shows that 
$ \lambda5007 / \rm H\beta > 5$ and $ \lambda6584/\rm H\alpha = 1.14$.

{\bf KUV 03079-0101} (Noguchi et al., 1980) is an emission line galaxy 
according to Chaffee et al. (1991). Our spectrum (Fig. 3) shows 
broad Balmer lines (3000 kms$^{-1}$ FHWM) and narrow 
$\lbrack \rm OIII \rbrack$ lines. The ratio 
of the total H$\beta$ flux to the $\lambda5007$ flux is R=10; this object 
is therefore a Seyfert 1.0 galaxy (Winkler, 1992).

\begin{figure*}[t]
\epsfxsize=18cm \epsfysize=15cm \epsfbox{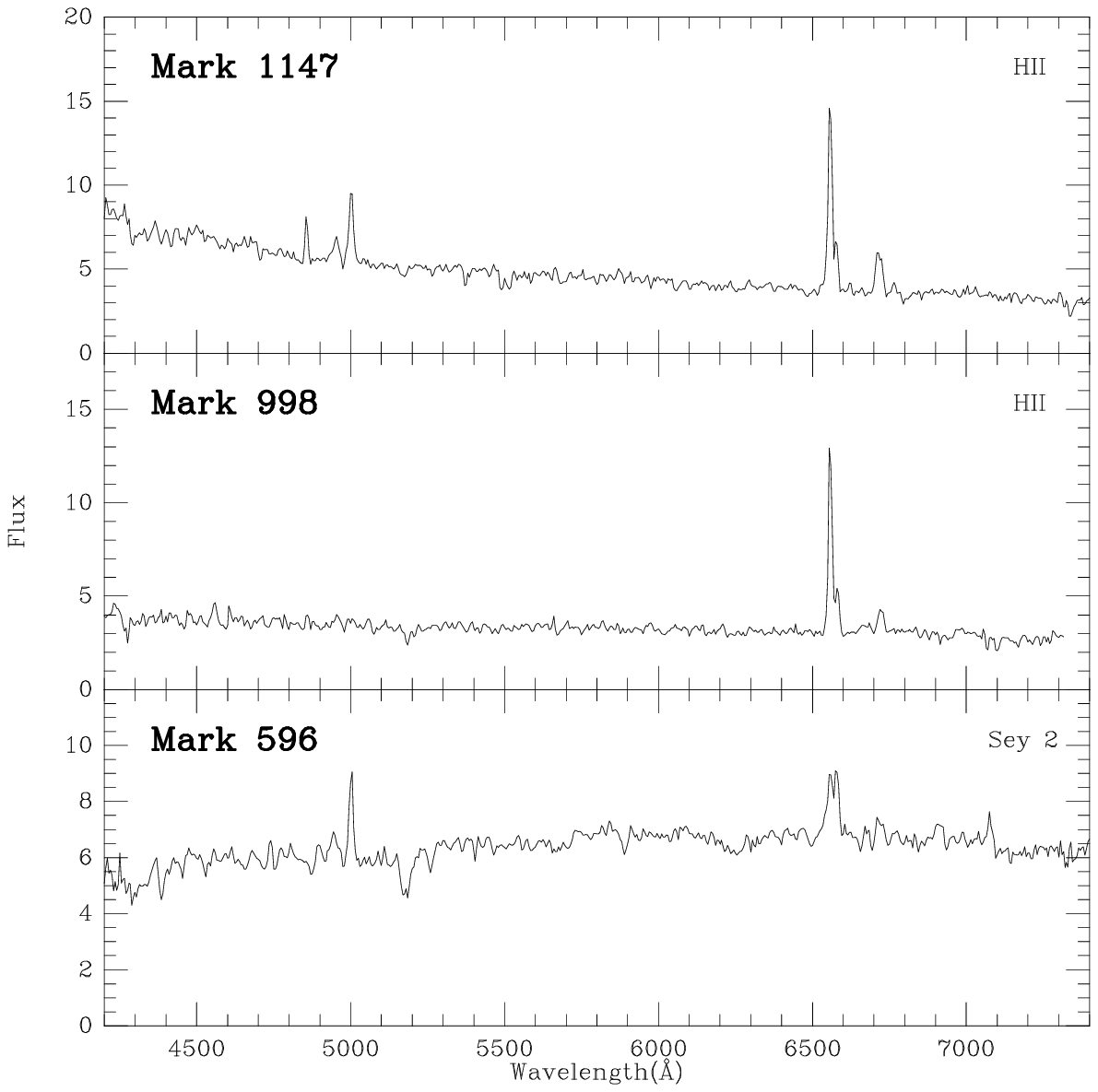}
\caption{Low dispersion spectra (resolution $\sim 13.5$ \AA), 
in the rest frame, of three objects
observed with the 1.93m OHP telescope. The fluxes are in units of
$\rm 10^{-16}\,erg\,s^{-1}cm^{-2}\AA^{-1}$.}
\label{LD_a}
\end{figure*}

{\bf CBS 74} is a Seyfert galaxy according to Wagner et al. 
(1988). It was not detected at 4850 MHz by Gregory \& Condon (1991) 
(S $< 25$ mJy) and is therefore a radioquiet object.
Our spectra (Fig. 3, 9) show that it is a Seyfert 1.2 galaxy with a 
very broad H$\alpha$ component (FWHM $\sim$ 12000 kms$^{-1}$) and R=3.6. 
Such broad lines are 
common in radioloud quasars and broad line radiogalaxies 
(Miley \& Miller, 1979; 
Wills \& Browne, 1986; Brotherton et al., 1994; Eracleous \& Halpern, 1994), 
but they are rare in Seyfert galaxies although a few cases are known, 
such as 2E 0450-1816 (Eracleous \& Halpern, 1994) and Arp 102B 
(Chen \& Halpern, 1989). Indeed, powerful radiogalaxies and 
radioloud quasars with extended radio morphologies tend to have the 
broadest Balmer lines, while AGNs with compact radiostructure and 
radioquiet objects have narrower Balmer lines 
(Miley \& Miller, 1979; Steiner, 1981; Wills \& Browne, 1986).

{\bf HS 0843+2533}. The ROSAT X-ray source RX J08469+2522 was identified 
by Bade et al. (1995) with a 16.8 mag AGN called HS 0843+2533, which exhibits 
a broad H$\alpha$ emission line (FWHM = 5900 kms$^{-1}$). Our 
spectrum (Fig. 9) shows, indeed, a strong, broad H$\alpha$ emission line 
(FWHM = 4850 kms$^{-1}$); this object is therefore a Seyfert 1 galaxy.

\begin{figure*}[t]
\epsfxsize=18cm \epsfysize=15cm \epsfbox{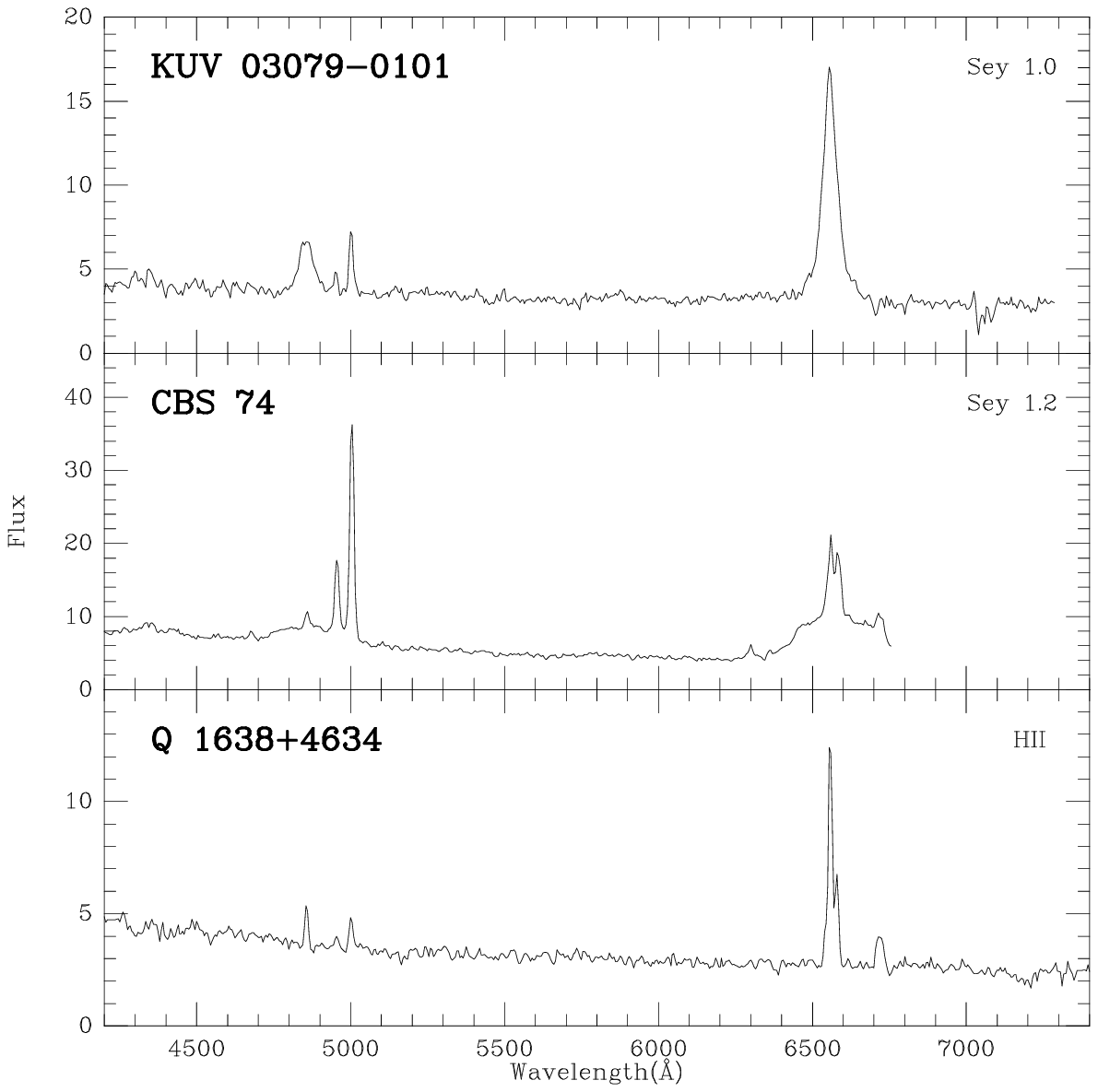}
\caption{Same as in Fig. 2 for three additional objects.}
\label{LD_2b}
\end{figure*}

{\bf Mark 391} = NGC 2691 is a S0a galaxy (Huchra, 1977). 
For Arakelian et al. (1972), it weakly 
shows the characteristics of the Seyfert nuclei, with a broad H$\alpha$ 
emission line ($\sim 50 \rm \AA$). On this basis, V\'eron-Cetty \& V\'eron 
(1985) classified it as a Seyfert 1. Shuder \& Osterbrock (1981), however, 
concluded from their own spectroscopic observations that it is not a Seyfert. 
Our spectra (Fig. 4) show Balmer lines in absorption together with narrow 
(FWHM $< 215$ kms$^{-1}$) emission lines with 
$\lambda5007/\rm H\beta= 1.21$ and $\lambda6584/\rm H\alpha = 0.55$, 
proving that it is a HII region.

{\bf KUG 0929+324} is a moderate excitation 
($\lambda5007/\rm H\beta$ $= 2.68$) 
emission line galaxy with an heliocentric radial velocity 
$\rm V=1500\pm70$ kms$^{-1}$ according to Augarde et al. (1994). Our spectrum 
(Fig. 7) shows that it is a HII region with narrow emission lines 
(FWHM $< 150$ kms$^{-1}$) and $ \lambda6584/\rm H\alpha = 0.10 $. We found 
the radial velocity to be $\rm V=4740$ kms$^{-1}$ (Augarde, 1995, private 
communication, gave $\rm V=4478$ kms$^{-1}$).

{\bf CG 49}. Salzer et al. (1995) published line intensity ratios for 
this object as follows: $ \lambda5007/\rm H\beta = 11.68 $ and 
$ \lambda6584/\rm H\alpha = 0.30 $. The $\lambda6584$ line is 
too strong for a HII region and too weak for a Seyfert 2 galaxy. Our 
spectrum shows $\lambda6584$ with the same low intensity, 
however it so happens that the redshifted wavelength of the 
$\lambda6584$ line (6874 \AA) falls precisely at the 
position of the atmospheric B band. When corrected for atmospheric absorption 
by dividing the observed spectrum by the spectrum of a standard star 
(Fig. 7), we 
obtain $ \lambda6584/\rm H\alpha = 0.79 $, a normal value for a Seyfert 2 
galaxy. The measured FWHM of the emission lines is $\sim 300$ kms$^{-1}$, 
in agreement with the adopted classification.

\begin{figure*}[t]
\epsfxsize=17.4cm \epsfysize=16.7cm \epsfbox{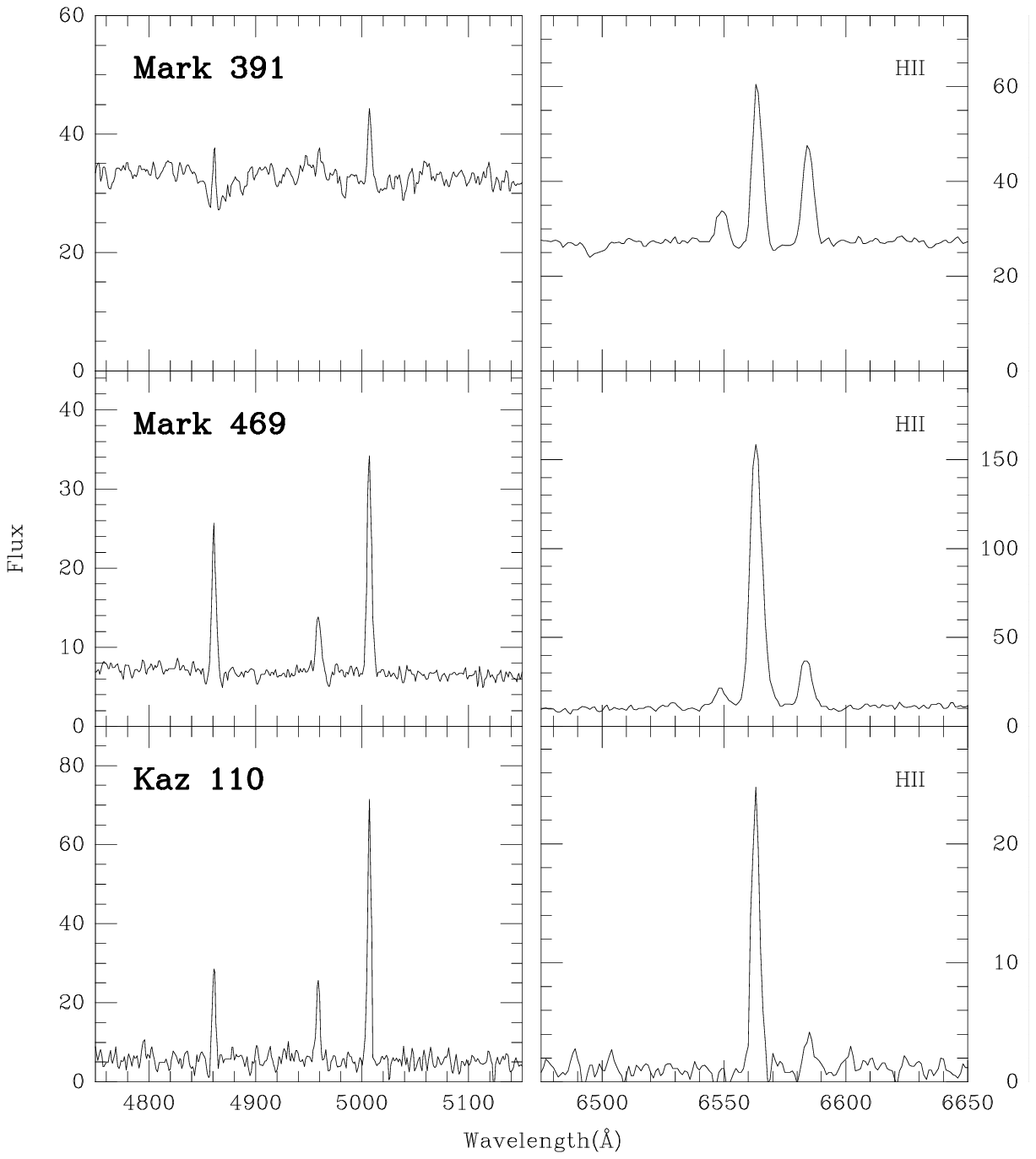}
\caption{Blue and red high dispersion spectra (resolution $\sim 3.4$ \AA), 
in the rest frame, of three objects observed with the 1.93m OHP telescope. 
The fluxes are in units of $\rm 10^{-16}\,erg\,s^{-1}cm^{-2}\AA^{-1}$.}
\label{BR_1}
\end{figure*}

\begin{figure*}[t]
\epsfxsize=17.4cm \epsfysize=16.7cm \epsfbox{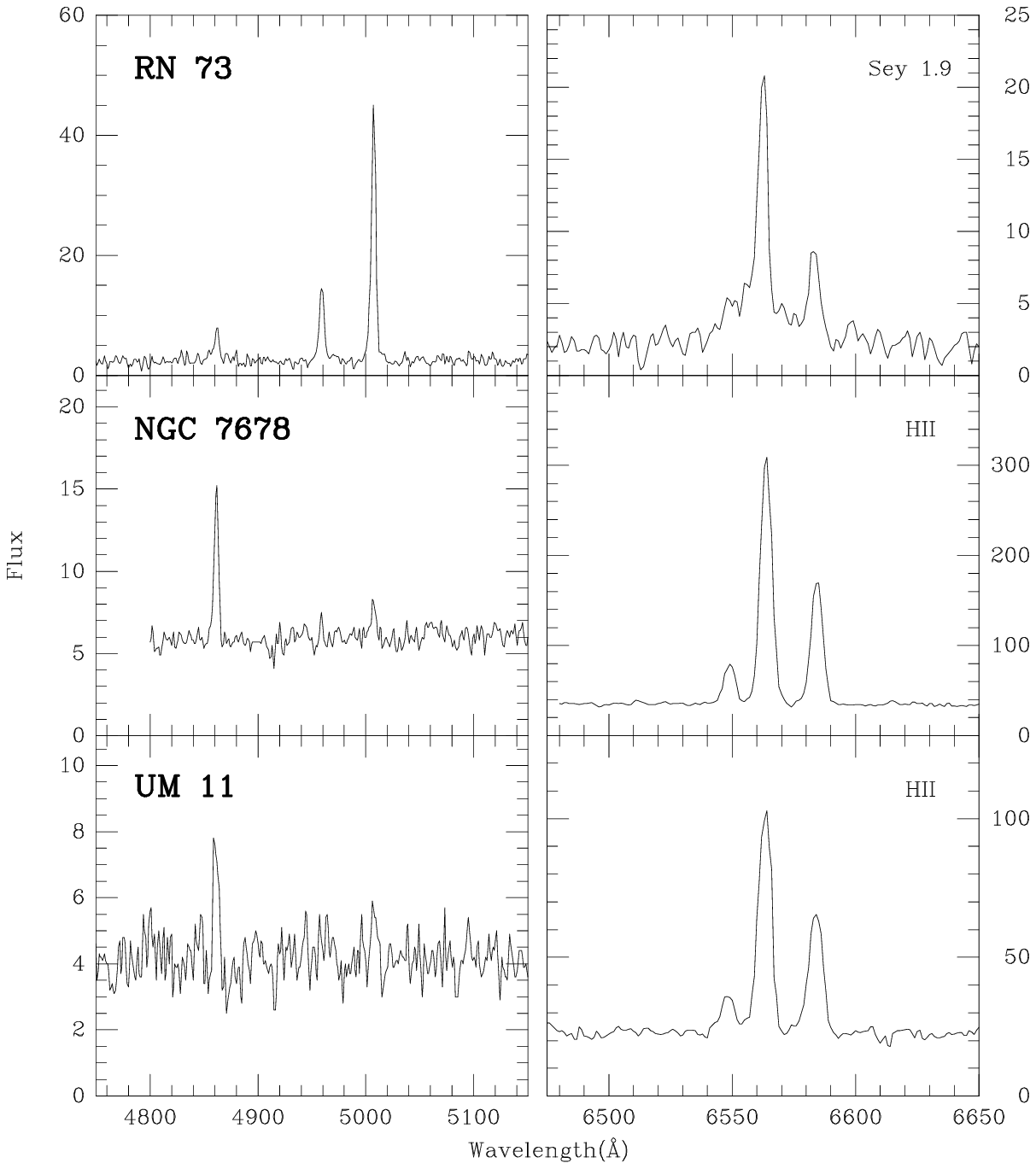}
\caption{Same as in Fig. 4 for three additional objects.}
\label{BR_2}
\end{figure*}

{\bf UM 446} is a moderate excitation ($\lambda5007/\rm H\beta = 4.54$) 
emission line galaxy (Salzer et al., 1989). Our spectrum (Fig. 7) shows 
narrow emission lines (FWHM $< 160$ kms$^{-1}$) with 
$ \lambda6584/\rm H\alpha = 0.04 $; this object is therefore a HII region.

{\bf US 2896} (Huang \& Usher, 1984) = CS 109 (Sandu\-leak \& Pesch, 1984) is 
an emission line galaxy (Mitchell et al., 1984), and a Seyfert 
1.5 galaxy according to Everett \& Wagner (1995). This is confirmed by our 
spectrum (Fig. 7) which shows a broad H$\alpha$ component 
(2100 kms$^{-1}$ FWHM). 
A $\lbrack \rm OI \rbrack \lambda6300$ emission line is observed with  
$\lambda6300 / \rm H\alpha_{narrow}$ = 0.03.

\begin{figure*}[t]
\epsfxsize=17.4cm \epsfysize=16.7cm \epsfbox{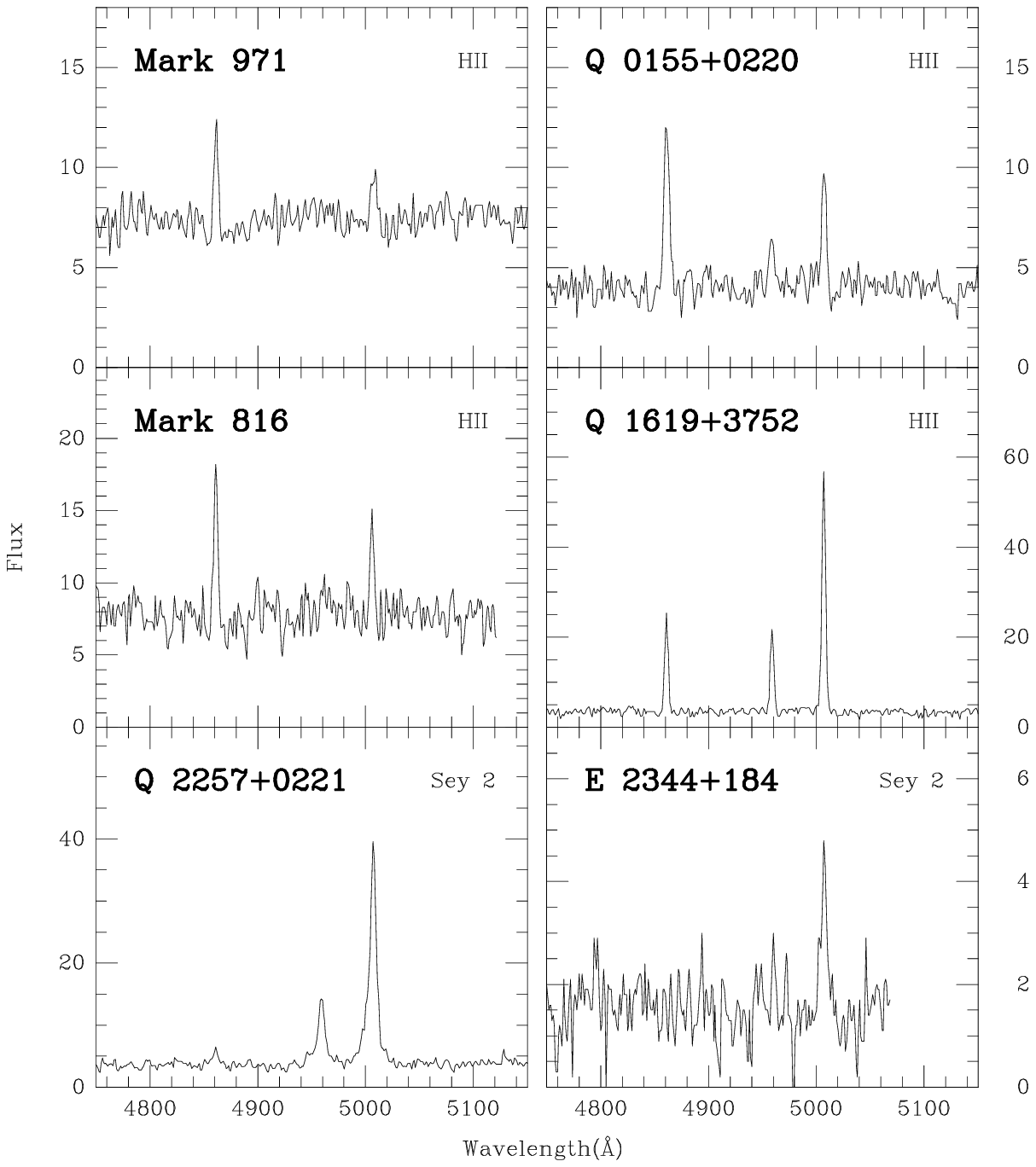}
\caption{Blue high dispersion spectra (resolution $\sim 3.4$ \AA), 
in the rest frame, for six objects observed with the 1.93m telescope. 
The fluxes are in units of
$\rm 10^{-16}\,erg\,s^{-1}cm^{-2}\AA^{-1}$.}
\label{Blue}
\end{figure*}

{\bf Mark 646}, PG 1203+35 (Green et al., 
1986), CG 885 (Pesch \& Sandu\-leak, 1988) or KUG 1203+354 (Takase \& 
Miyauchi-Isobe, 1991a) is a Seyfert galaxy according to Green et al. (1986).
Our spectrum (Fig. 7) shows a broad H$\alpha$ component 
(2350 kms$^{-1}$ FWHM); 
Mark 646 is therefore a Seyfert 1 galaxy. A $\lambda6300$ emission line 
is observed with $\lambda6300 / \rm H\alpha_{narrow}$ = 0.05.

{\bf 2E 1219+0447} is an emission line galaxy (Bothun et al., 
1984; Margon et al., 1985). We classify it as a Seyfert 1 galaxy on the basis 
of a weak, broad (FWHM $\sim$ 8500 kms$^{-1}$) H$\alpha$ component (Fig. 9).

{\bf KUV 13000+2908} (Noguchi et al., 1980), CG 963 (Sanduleak \& Pesch, 
1990) or PB 3241 (Berger et al., 1991) is a Seyfert 2 galaxy according 
to Wegner \& McMahan (1988). Our spectrum (Fig. 7) shows narrow 
(FWHM $< 185$ kms$^{-1}$) emission lines with 
$ \lambda6584/\rm H\alpha < 0.1$, so this object is a HII region. 

\begin{figure*}[t]
\epsfxsize=17.4cm \epsfysize=16.7cm \epsfbox{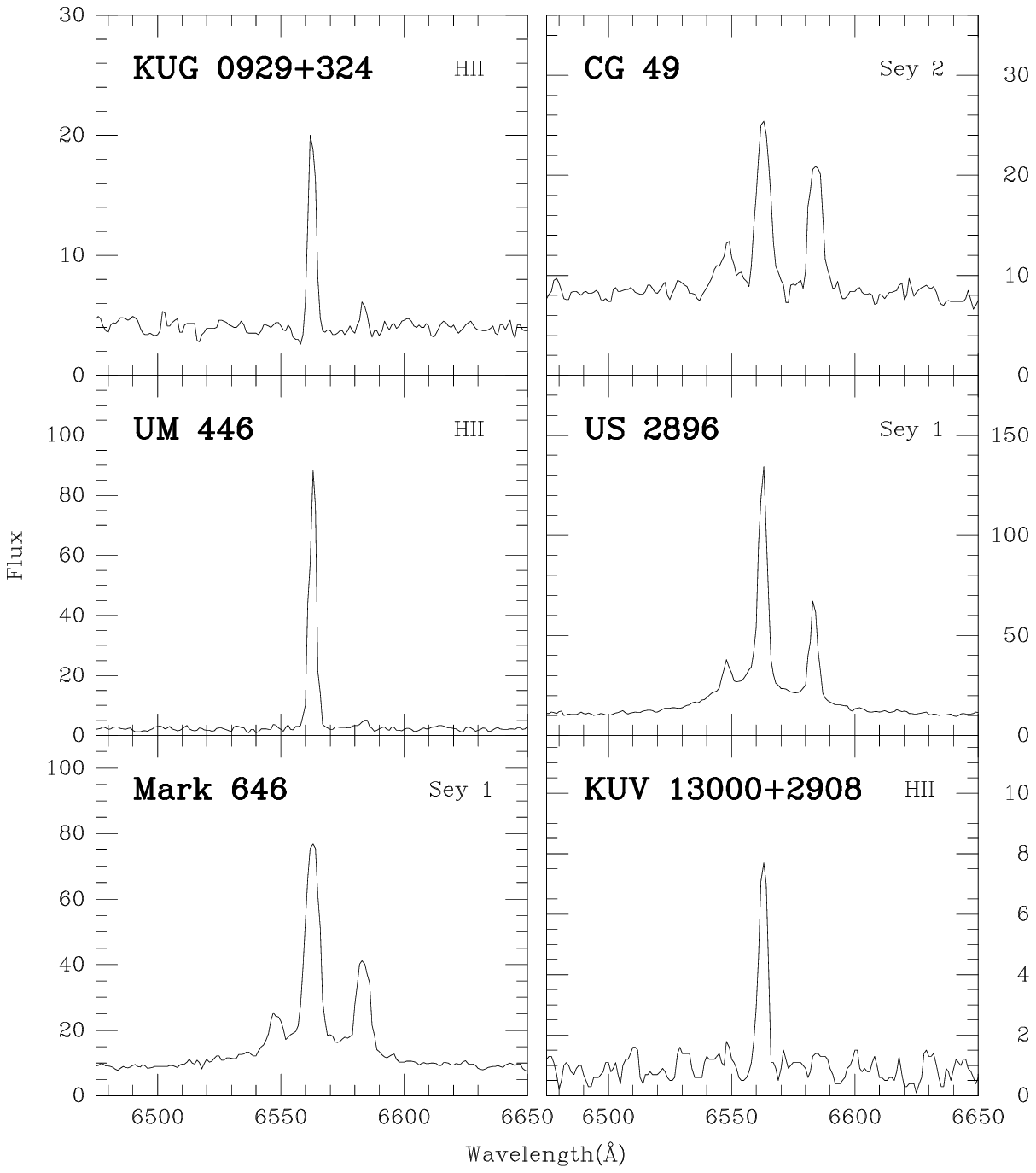}
\caption{Red high dispersion spectra 
(resolution $\sim 3.4$ \AA), in the rest frame, for six
objects observed with the 1.93m telescope. The fluxes are in units of
$\rm 10^{-16}\,erg\,s^{-1}cm^{-2}\AA^{-1}$.}
\label{Red_1}
\end{figure*}

{\bf Q 1356-067} is a QSO according to Goldschmidt et al. 
(1992). Our spectrum (Fig. 8), however, shows narrow (FWHM $< 280$ kms$^{-1}$) 
emission lines with $ \lambda6584/\rm H\alpha = 0.16 $; this object is 
therefore a HII region. A $\lambda6300$ emission line 
is observed with $\lambda6300 / \rm H\alpha = 0.015$

{\bf Mark 469}, CG 899 (Pesch \& Sanduleak, 1989) or KUG 1416+345 (Takase \& 
Miyauchi-Isobe, 1984) is a 16.1 mag, UV excess galaxy (Peterson et al., 1981). 
Our observations (Fig. 4) show that 
it is a HII region, with 
$ \lambda5007/\rm H\beta = 1.38 $ and $ \lambda6584/\rm H\alpha = 0.18 $, 
the lines being narrow (FWHM $< 260$ kms$^{-1}$). 

{\bf PKS 1420-27}. This radiosource was identified by Bolton \& Ekers (1966) 
with an 18 mag QSO. The identification was later confirmed by accurate 
optical and radio position measurements (Hunstead 1971, 1972). Our spectrum 
(Fig. 1) shows that it is indeed a QSO at $z=0.985$.
The emission
line fluxes are 83 and 53 $10^{-16}\, \rm erg\,s^{-1}\,cm^{-2}$ for 
$\rm CIII \rbrack\lambda1909$ and $\rm MgII\,\lambda2798$ respectively. 

{\bf Mark 816} = KUG 1431+529 (Takase \& Miyauchi-Isobe, 1985) is a 16.5 
mag, possibly Seyfert, galaxy (Afanasev et al., 1979); however, 
$ \lambda6584/\rm H\alpha < 0.3$ (Afanasev et al., 1980). Our spectrum 
(Fig. 6) shows narrow emission lines ($< 240$ kms$^{-1}$ FWHM) with 
$ \lambda5007/\rm H\beta = 0.63 $. This object is most 
probably a HII region. 
  
{\bf PKS 1437-153} is a flat spectrum radiosource identified by Condon et al. 
(1977) with a 19.0 mag starlike object. It has a featureless spectrum 
between 3800 and 7000 \AA\, (Fig. 1) and is most probably a BL Lac object.

\begin{figure*}[t]
\epsfxsize=17.4cm \epsfysize=16.7cm \epsfbox{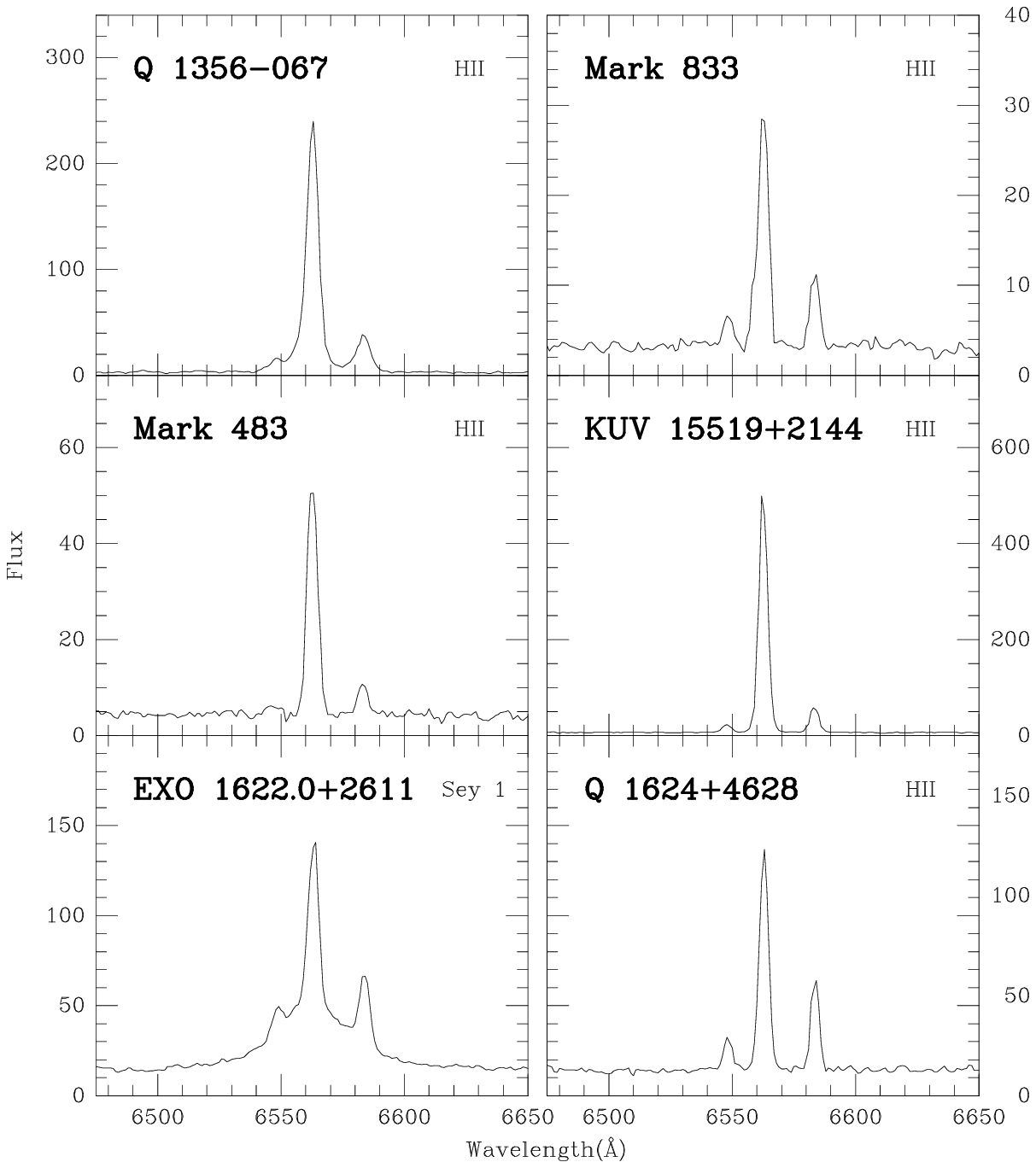}
\caption{Same as in Fig. 7 for six additional objects.}
\label{Red_2}
\end{figure*}

{\bf Mark 833} = CG 590 (Sanduleak \& Pesch, 1987) is an emission line 
galaxy (Markarian et al., 1985) which has been called a "narrow-line active 
galactic nucleus" by  Veilleux \& Osterbrock (1987), based on emission-line
intensity ratios published by Shuder \& Osterbrock (1981); it however 
happened that the object studied in this last paper is Mark 883 which is, in a
few occasions mistakenly called Mark 833 (H. Falcke, private communication).
The nature of the emission-line nebulosity in Mark 833 was therefore unknown.  
Our spectrum (Fig. 8) shows narrow (FWHM $< 225$ kms$^{-1}$) 
emission lines 
with $ \lambda6584/\rm H\alpha = 0.30 $. This object is, therefore, 
a HII region.

{\bf Mark 483} = CG 741 (Sanduleak \& Pesch, 1987) is 
an emission line galaxy (Markarian et al., 1988; Izotov et al., 
1993) with a strong UV excess (U-B $=-0.45$, Peterson et al., 1981). The 
emission-line ratios published by Markarian et al. 
($ \lambda6584/\rm H\alpha < 0.33 $) and Izotov et al. 
($ \lambda5007/\rm H\beta \sim 2.0 $) suggested that it is a HII region. This 
is confirmed by our spectrum (Fig. 8) which shows narrow emission lines 
(FWHM $< 225$ kms$^{-1}$) with $ \lambda6584/\rm H\alpha =0.12 $.

{\bf KUV 15519+2144} is a Seyfert 2 galaxy according to Wagner \& 
Swanson (1990). Our spectrum (Fig. 8) shows it to be a HII region, 
with narrow 
(FWHM $< 185$ kms$^{-1}$) emission lines and 
$ \lambda6584/\rm H\alpha = 0.11 $. A $\lambda6300$ emission line is 
observed with $\lambda6300 / \rm H\alpha = 0.02$
 
{\bf Q 1619+3752}, an emission line galaxy according to 
Schneider et al. (1994), is classified as a HII region, as it shows 
narrow emission lines 
($< 240$ kms$^{-1}$ FWHM) and $ \lambda5007/\rm H\beta = 2.56$ (Fig. 6).

{\bf EXO 1622.0+2611}. An AGN for Giommi et al. (1991), this is a Seyfert 1 
galaxy, as it presents a broad H$\alpha$ component (1770 kms$^{-1}$ 
FWHM) (Fig. 8).

{\bf Q 1624+4628}. An emission line galaxy according to Schneider 
et al. (1994), it is a HII region, with narrow ($< 195$ kms$^{-1}$ FWHM) 
emission lines and $ \lambda6584/\rm H\alpha = 0.40 $ (Fig. 8).

{\bf Q 1638+4634}. An emission line galaxy according to Schneider 
et al. (1994), it is a HII region, with $ \lambda5007/\rm H\beta = 0.86 $ 
and $ \lambda6584/\rm H\alpha = 0.40 $ (Fig. 3).

{\bf Kaz 110}. The emission-line gas in this object was shown to be 
ionized by hot stars (Kazarian \& Tamazian, 1993). Our spectra (Fig. 4) 
confirm this result, the measured line ratios being: 
$ \lambda5007/\rm H\beta = 2.83 $ 
and $ \lambda6584/\rm H\alpha = 0.09 $.

\begin{figure*}[t]
\epsfxsize=17.4cm \epsfysize=12.2cm \epsfbox{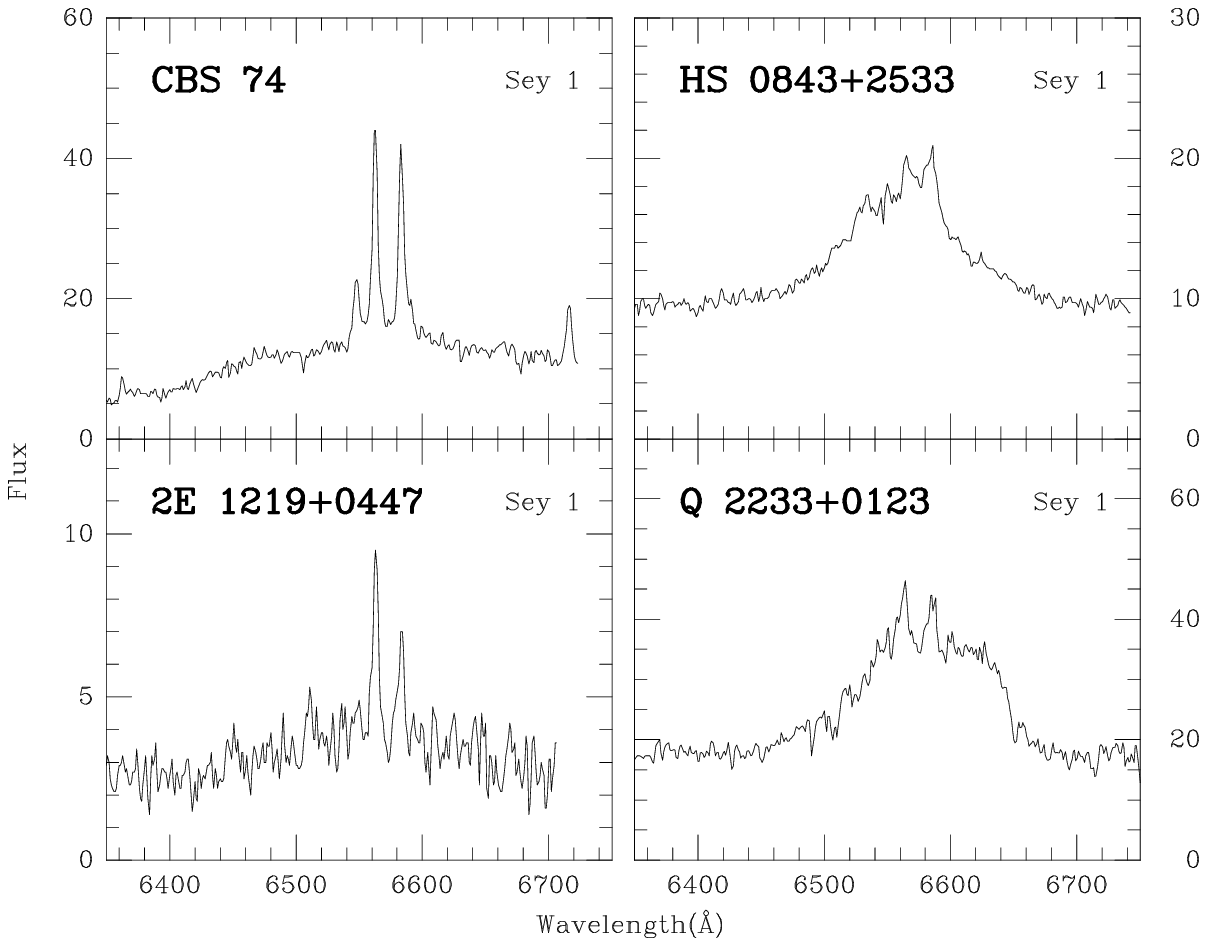}
\caption{Same as in fig. 7 for four additional objects.}
\label{Red_3}
\end{figure*}

{\bf PKS 1903-80}. This flat spectrum radiosource (Quiniento \& Cersosimo, 
1993) was identified with a 19.0 mag QSO by Anguita et al. (1979). The 
identification was confirmed by an accurate radioposition measurement 
(Russel et al., 1992). It is indeed a QSO at $z=1.756$ (Fig. 1).
The emission
line fluxes are 57 and 25 $10^{-16}\, \rm erg\,s^{-1}\,cm^{-2}$ for 
$\rm CIV \lambda1550$ and $\rm CIII \rbrack\lambda1909$ res\-pectively. 

{\bf RN 73} (Ryle \& Neville, 1962) = 8C 2037+880 (Rees, 1990) 
was identified with a 17.5 mag emission line galaxy (Penston, 1971). 
Our spectra (Fig. 5) show a weak, broad 
(FWHM $\sim 1590$ kms$^{-1}$) H$\alpha$ component, but no broad H$\beta$ 
component: this object is a Seyfert 1.9 galaxy. However, the ratio 
$\lambda6584/\rm H\alpha_{narrow} = 0.37$ is low for a Seyfert galaxy. 
Halliday (1977) published an accurate radiomap for this source; its position, 
as measured on this map ($\alpha_{1950} = \rm 20^{h}36^{m}44^{s}$, 
$\delta_{1950} = 88^{\rm o}01'58''$), is about 20 arcsec away from the 
position of the galaxy, suggesting that the radiostructure and the galaxy 
are not to be related.

{\bf Q 2233+0123}. An emission line galaxy according to 
Schneider et al. (1994), it is a Seyfert 1, having a strong and 
broad (FWHM $\sim 5500$ kms$^{-1}$) H$\alpha$ component (Fig. 9). 
The profile of this 
line deviates significantly from a Gaussian, having a flat top.

{\bf Q 2257+0221} is an emission line galaxy according to 
Schneider et al. (1994). It is a Seyfert 2 having broad 
(FWHM $\sim 500$ kms$^{-1}$), asymmetrical 
$\lbrack \rm OIII \rbrack$ lines, that are much stronger than 
H$\beta$ ($ \lambda5007/\rm H\beta \sim 17 $) (Fig. 6).

\begin{table*}[t]
\begin{center}
\begin{tabular}{p{2.6cm}crrrrrrrrc}
\hline
Name & z & $\rm V\;\;\;\; $ & FWHM\/ & \underline{$\lambda5007\:$} 
& H$\beta\; $ & $\rm V\;\;\;\; $& FHWM\/ & \underline{$\lambda6584\:$} 
& H$\alpha\; $ & Spectral \\
 & & $\;\;\/ \rm (kms^{-1})$ & $\rm (kms^{-1})$ & H$\beta\; $ & 
& $\;\;\/ \rm (kms^{-1})$ & $\rm (kms^{-1})$ & H$\alpha\;\; $ & 
& type \\
\hline
Mark 1147 & 0.0364 & 14\verb+  + & (780)\verb+ + & 2.18 & 2.3 
& 11\verb+  + & (635)\verb+ + & 0.24 & 11.4 & HII \\
Mark 971 & 0.0823 & 15\verb+  + & 280\verb+ + & 0.41 & 5.6 
& --\verb+  + & --\verb+  + & --\verb+  + & --\verb+ + & HII \\
Mark 998 & 0.0761 & --\verb+  + & --\verb+  + & --\verb+  + & --\verb+ + 
& 22\verb+  + & (580)\verb+ + & 0.23 & 10.3 & HII \\
Q $0155+0220$ & 0.0651 & -1\verb+  + & 325\verb+ + & 0.71 & 8.4 & --\verb+  + 
& --\verb+  + & --\verb+  + & --\verb+ + & HII \\
Mark 596 & 0.0388 & -16\verb+  + & (960)\verb+ + & $>5.0$\verb+ + 
& $<0.6$ & 45\verb+  + & (720)\verb+ + & 1.14 & 2.2 & S2 \\
KUV $03079-0101$ & 0.0807 & -15\verb+  + & (860)\verb+ + 
& $>10.0$\verb+ + & $<0.4$ 
& --\verb+  + & --\verb+  + & --\verb+  + & --\verb+ + & S1.0 \\
 & & -29\verb+  + & 3360\verb+ + & --\verb+  + 
& 3.1 & -28\verb+  + & 2570\verb+ + & --\verb+  + & 12.2 &  \\
CBS 74 & 0.0920 & 143\verb+  + & (945)\verb+ + & 15.6\verb+ + & 1.9 & -10\verb+  + 
& 260\verb+ + & 0.89 & 27.8 & S1.2 \\
 & & 1485\verb+  + & 14500\verb+ + & --\verb+  + & 2.3 & 1092\verb+  + 
& 12200\verb+ + & --\verb+  + & 10.4 & \\
HS 0843+2533 & 0.0507 & --\verb+  + & --\verb+  + & --\verb+  + & --\verb+ + 
& 90\verb+  + & 475\verb+ + & 1.6\verb+ + & 2.5 & S1 \\
 & & --\verb+  + & --\verb+  + & --\verb+  + & --\verb+ + & -70\verb+  + 
& 4850\verb+ + & --\verb+  + & 7.3 & \\
Mark 391 & 0.0133 & 9\verb+  + & 210\verb+ + & 1.21 & 9.9 & 45\verb+  + 
& 225\verb+ + & 0.55 & 36.7 & HII \\
KUG 0929+324 & 0.0158 & --\verb+  + & --\verb+  + & --\verb+  + & --\verb+ + 
& -6\verb+  + & 150\verb+ + & 0.10 & 17.6 & HII \\
CG 49 & 0.0438 & --\verb+  + & --\verb+  + & --\verb+  + & --\verb+ + 
& 15\verb+  + & 300\verb+ + & 0.79 & 17.4 & S2 \\
UM 446 & 0.0061 & --\verb+  + & --\verb+  + & --\verb+  + & --\verb+ + 
& 9\verb+  + & 160\verb+ + & 0.04 & 79.8 & HII \\
US 2896 & 0.0594 & --\verb+  + & --\verb+  + & --\verb+  + & --\verb+ + 
& -3\verb+  + & 185\verb+ + & 0.43 & $103\verb+ +\;$ & S1 \\
 & & --\verb+  + & --\verb+  + & --\verb+  + & --\verb+ + & -120\verb+  + 
& 2100\verb+ + & --\verb+  + & 12.3 & \\
Mark 646 & 0.0536 & --\verb+  + & --\verb+  + & --\verb+  + & --\verb+ + 
& 0\verb+  + & 280\verb+ + & 0.46 & 62.8 & S1 \\
 & & --\verb+  + & --\verb+  + & --\verb+  + & --\verb+ + 
& -16\verb+  + & 2350\verb+ + & --\verb+  + & 8.0 &  \\
2E 1219+0447 & 0.0947 & --\verb+  + & --\verb+  + & --\verb+  + & --\verb+ + 
& 3\verb+  + & 225\verb+ + & 0.54 & 5.5 & S1 \\
 & & --\verb+  + & --\verb+  + & --\verb+  + & --\verb+ + & 222\verb+  + 
& 8440\verb+ + & --\verb+  + & 1.6 & \\
KUV 13000+2908 & 0.0223 & --\verb+  + & --\verb+  + & --\verb+  + & --\verb+ + 
& 3\verb+  + & 185\verb+ + & $<0.1$\verb+ + & 7.0 & HII \\
Q 1356-067 & 0.0746 & --\verb+  + & --\verb+  + & --\verb+  + & --\verb+ + 
& 6\verb+  + & 280\verb+ + & 0.16 & $226\verb+ +\,$ & HII \\
Mark 469 & 0.0689 & -9\verb+  + & 295\verb+ + & 1.38 & 19.4 & 18\verb+  + 
& 260\verb+ + & 0.18 & $147\verb+ +\;$ & HII \\
Mark 816 & 0.0887 & -19\verb+  + & 240\verb+ + & 0.63 & 10.5 & --\verb+  + & 
--\verb+  + & --\verb+  + & --\verb+ + & HII \\
Mark 833 & 0.0395 & --\verb+  + & --\verb+  + & --\verb+  + & --\verb+ + 
& -8\verb+  + & 225\verb+ + & 0.30 & 25.7 & HII \\
Mark 483 & 0.0481 & --\verb+  + & --\verb+  + & --\verb+  + & --\verb+ + 
& -5\verb+  + & 225\verb+ + & 0.12 & 48.3 & HII \\
KUV 15519+2144 & 0.0392 & --\verb+  + & --\verb+  + & --\verb+  + & --\verb+ + 
& -9\verb+  + & 185\verb+ + & 0.11 & $483\verb+ +\;$ & HII \\
Q 1619+3752 & 0.0331 & -3\verb+  + & 240\verb+ + & 2.56 & 20.4 & --\verb+  + & 
--\verb+  + & --\verb+  + & --\verb+ + & HII \\
EXO 1622.0+2611 & 0.0394 & --\verb+  + & --\verb+  + & --\verb+  + & --\verb+ + 
& 15\verb+  + & 225\verb+ + & 0.40 & 93.0 & S1 \\
 & & --\verb+  + & --\verb+  + & --\verb+  + & --\verb+ + & 12\verb+  + 
& 1770\verb+ + & --\verb+  + & 33.0 & \\
Q 1624+4628 & 0.0301 & --\verb+  + & --\verb+  + & --\verb+  + & --\verb+ + 
& 6\verb+  + & 195\verb+ + & 0.40 & $124\verb+ +\;$ & HII \\
Q 1638+4634 & 0.0581 & 10\verb+  + & (720)\verb+ + & 0.86 & 1.8 & 36\verb+  + 
& (625)\verb+ + & 0.40 & 10.3 & HII \\
Kaz 110 & 0.0527 & -5\verb+  + & 225\verb+ + & 2.83 & 22.9 
& 9\verb+  + & 170\verb+ + & 0.09 & 23.1 & HII \\
RN 73 & 0.0491 & 18\verb+  + & 310\verb+ + & 7.4\verb+ + & 5.4 
& -16\verb+  + & 185\verb+ + & 0.37 & 16.3 & S1.9 \\
 & & --\verb+  + & --\verb+  + & --\verb+  + & --\verb+ + 
& -34\verb+  + & 1590\verb+ + & --\verb+  + & 3.1 & \\
Q 2233+0123 & 0.0566 & --\verb+  + & --\verb+  + & --\verb+  + & --\verb+ + 
& 15\verb+  + & 290\verb+ + & 0.60 & 8.5 & S1 \\
 & & --\verb+  + & --\verb+  + & --\verb+  + & --\verb+ + 
& 1080\verb+  + & 5500\verb+ + & --\verb+  + & 21.3 & \\
Q 2257+0221 & 0.0466 & 3\verb+  + & 495\verb+ + & 16.7\verb+ + 
& 1.9 & --\verb+  + & --\verb+  + & --\verb+  + & --\verb+ + & S2 \\
NGC 7678 & 0.0116 & 12\verb+  + & 270\verb+ + & 0.25 & 9.5 
& 45\verb+  + & 250\verb+ + & 0.52 & $272\verb+ +\;$ & HII \\
E 2344+184 & 0.1365 & 24\verb+  + & 425\verb+ + & $> 6.0$\verb+ + 
& $< 0.5$ & --\verb+  + & --\verb+  + & --\verb+  + & --\verb+ + & S2\\
UM 11 & 0.0390 & -8\verb+  + & 310\verb+ + & 0.46 & 3.8 
& 21\verb+  + & 290\verb+ + & 0.56 & 80.2 & HII \\
\hline
\end{tabular}
\caption{Fitting profile analysis results (see text). Col. 2 gives the 
redshift used to deredshift the spectra, while col. 3 and 7 give the 
velocity of the lines as measured on the dereshifted spectra. Col. 6 and 10 
give the peak intensity of the H$\beta$ and H$\alpha$ lines respectively 
(in units of $10^{-16}\, \rm erg\,s^{-1}\,cm^{-2}\,\AA^{-1}$). The FWHM are 
observed values, not corrected for the instrumental profile; values in 
parenthesis are from low dispersion spectra and are unresolved.} 
\end{center}
\end{table*}

{\bf NGC 7678} = Kaz 336 (Kazarian \& Kazarian, 1980). Although classified as 
a Seyfert 2 galaxy by Kazarian (1993), this is a HII region 
with $\lambda6584/\rm H\alpha = 0.52$, $\lambda5007/\rm H\beta = 0.25 $ 
and linewidth $< 260$ kms$^{-1}$ FWHM (Fig. 5).

{\bf E 2344+184} is a spiral galaxy (Hutchings \& Neff, 1992). 
According to Margon et al. (1985), it is an emission line 
galaxy with a strong $\lambda6584$ 
emission line. Our low signal-to-noise blue spectrum (Fig. 6) shows a strong 
$\lambda5007$ line ($\lambda5007/\rm H\beta > 6$), 
so it is most probably a Seyfert 2 galaxy.

{\bf UM 11}. Terlevich et al. (1991) gave line ratios: 
$\lambda6584/\rm H\alpha = 1.23 $ and $\lambda5007/\rm H\beta = 0.83 $, 
suggesting that this object is a Liner; our spectra (Fig. 5) give 
$\lambda6584/\rm H\alpha = 0.56$ and $\lambda5007/\rm H\beta \sim 0.5$ 
(with linewidth $< 300$ kms$^{-1}$ FWHM) showing that it is a 
HII region instead.

\section{Conclusions}
We have observed 37 AGN candidates and classified them on the basis of their 
spectroscopic properties; the line intensities and widths were obtained 
by fitting the spectra with Gaussian components. 
We concluded that three of the observed objects are confirmed 
QSOs, one is a BL Lac object, 
nine are Seyfert 1 galaxies, four are Seyfert 2s, while twenty are 
HII regions.\\

{\it Acknowledgements.} We are grateful to Dr. M. Viton for having allowed us 
to take the low dispersion spectrum of CBS 74 during his observing run. 
This research has made use of the NASA/IPAC 
extragalactic database (NED) which is operated by the Jet Propulsion 
Laboratory, Caltech, under contract with the National Aeronautics and Space 
Administration. A. C. Gon\c{c}alves acknowledges support from the JNICT
during the course of this work, in the form of a PRAXIS XXI PhD. grant
(PRAXIS XXI/BD/5117/95).


\end{document}